# The Problem of Modeling of Economic Dynamics

S. I. Chernyshov,[1,2] A. V. Voronin,[2*] and S. A. Razumovsky[3**]

[1]*Kharkov National Economic University (Kharkov, Ukraine)*
[2]*"Lemma" Insurance Company (Kharkov, Ukraine)*
[3]*Karkov State Academy of Physical Culture (Kharkov, Ukraine)*
[*]E-mail address: voronin61@ukr.net
[**]E-mail address: rsa_777@mail.ru
(in Russian: http://chvr-article.narod.ru)

**Abstract**

The correctness of Harrod's model in the differential form is studied. The inadequacy of exponential growth of economy is shown; an alternative result is obtained. By example of Phillips' model, an approach to correction of macroeconomic models (in terms of initial prerequisites) is generalized. A methodology based on balance relations for modelling of economic dynamics, including obtaining forecast estimates, is developed. The problems thus considered are reduced to the solution of Volterra and Fredholm integral equations of the second kind.

## Introduction

Originally, the authors' objective was to analyze the procedure of the construction of differential equations employed in modeling of macroeconomic processes. The results proved to be substantially unexpected, because a number of contradictions were found. In the course of investigations intended to resolve these contradictions, alternative concepts of mathematical modeling of economic processes were formed.

Section 1 of the paper is concerned with establishing incorrectness of Harrod's well-known model of economic growth. This incorrectness is caused by a procedure of the derivation of the differential equation using dependence whose character is essentially discrete. Accordingly, a contradiction emerges whose root is in an inadequate interpretation of the notion of infinitesimal. Regarding this issue, we present arguments that are based on fundamentals of the theory of generalized functions.

Solutions obtained when considering the same model in finite difference interpretation radically contradict exponential growth of the economy that is represented in a number of publications as an immediate consequence of Harrod's differential model.

In a constructive aspect, we propose an approach to improve on the initial relations in such a way that a model based on them be adequate to criteria of the applicability of continuous analysis. A solution is obtained that shows that the forecast of economic development can be made only for a limited period of time.

In section 2, we show that Harrod-Domar's model, formally represented in terms of intensities of flows, is, in reality, discrete. We also demonstrate the incorrectness of Phillips' model: this model has acquired the status of a classical one and is widely represented in special literature, including manuals. A method of correcting this model is proposed, which is based on the above-mentioned approach to the construction of relations between macroeconomic functions of different dimensions: the initial background of economic content is left unchanged at that.

The use of Phillips' "new" model is reduced to the solution of an ordinary differential equation of the second order whose coefficients, in contrast to the "classical" interpretation, are variables. This point substantially widens the spectrum of potentially possible ways of behavior of the economic system, and it seems to be important for practical applications. We give references to literature where versions of analytical solution of the posed problem can be found. A corresponding numerical algorithm is also given.

Arguments that relations of the balance of financial flows are the best fit to the objectives of economic-mathematical modelling form the ideological basis of the content of sections 3 and 4. At the beginning, by an iteration procedure, the static model of value balance is given time dependence, after which the use of a Taylor expansion allows us to derive a system of differential equations. After that, the Cauchy problem is reduced to a system of Volterra integral equations of the second kind: this system has rather favourable, from the point of view of numerical realization, properties.

In this regard, the "balance" is considered as an alternative to unjustified refraction in the economic sphere of the methodology of the construction of mathematical models taken from the field of natural sciences, such as mechanics. As a matter of fact, differential equations of this science organically follow from conditions of equilibrium of an infinitesimal element, whereas analogous constructions in problems of economics have no objective meaning.

Finally, in section 4, we develop an approach to the forecasting of the behavior of an economic system that includes a certain number of participants. The problem is reduced to the solution of the Fredholm integral equation of the second kind whose kernel exclusively depends on factors that characterize relations between the participants. In addition, the free term reflects the index of the cost price of production and planned results of the activity. We propose to refine on the forecast by means of carrying out variational evaluation in the range of changes in the cost price and the results for a resolvent less affected by dynamics.

As tools of such stabilization, there appear mechanisms of an effective interaction between the participants that are objectively inherent in an economic cluster. Here, we employ reasoning related to peculiarities of the solution to the integral equation of the second kind whose kernel depends on the argument, as well as techniques of matrix analysis.

**1. Harrod's model and macroeconomic growth**

Harrod's model of the development of the economy, in the representation of L. V. Kantorovich and A. B. Gorstko [1, pp. 160-161],[#] is defined by the following relations:

$$Y(t) = C(t) + S(t); \; S(t) = I(t); \; S(t) = mY(t);$$
$$d_t K(t) = I(t); \; K(t) = nY(t), \tag{1.1}$$

---

[#] Note that all page references in the text are given according to Russian editions of corresponding literature sources.

where $Y(t)$ is the national income; $C(t)$, $S(t)$ are volumes of consumption and accumulations per year; $I(t)$ a volume of investment per year; $K(t)$ is the capital. All these quantities are measured in terms of money equivalent; $d_t = d/dt$, and $t$ is dimensionless time; $0 \leq m \leq 1$ and $n$ are dimensionless constants. As regards $n$, it is characterized as the number of years during which the income counterbalances the capital. The differential equation that formally follows from (1.1) and the solution to this equation take, respectively, the form

$$d_t Y(t) = (m/n) Y(t), \ Y(t) = Y_0 e^{mt/n}, \ Y_0 = Y(0). \qquad (1.2)$$

However, we face a contradiction since in considering the income $Y(t)$ during infinitely small time interval, in the relationship (1.1) parameter $n \to \infty$ and the uncertainty appear. This must be taken into account when the construction of the differential equation is considered. Generally, as a result of the presence of the derivative $d_t K(t)$ in (1.1), obviously, the most suitable categories for the considered model are those of continuous analysis. From this point of view, it would be natural to interpret $Y(t)$, $C(t)$, $S(t)$ as $I(t)$ intensities of financial flows with respect to time.

However, in this case, mutual dependence between the function $K(t)$ and $Y(t)$ in (1.1) is of essentially discrete character. Therefore, instead of this dependence, adhering to the category of intensities of flows, it is logical to use the following model of the formation of capital:

$$K(t) = K_0 + \frac{n}{t} \int_e^t Y(h) dh, \ K_0 = K(0), \ 0 < e < t,$$

where $e$ is a small quantity that, in its continuous interpretation, practically realizes the prerequisite (1.1) of counterbalancing $K(t)$ by the income.

In other words, quantities

$$\int_e^t Y(h) dh; \ \frac{n}{t-e} \int_e^t Y(h) dh$$

are, respectively, the income during the interval ε to τ and the same income during the interval normalized to parameter $n$ (e.g. one year if we follow [1]). Actually, the proportion is used: let $n = 10$ for one year, then for the interval with duration $t$ we have $10/t$.

Seemingly,

$$d_t K(t) = -\frac{n}{t^2}\int_e^t Y(h)dh + \frac{n}{t}Y(t),$$

and, eliminating the derivative by means of the expression $d_t K(t) = mY(t)$ that follows from (1.1), we obtain, as a result of comparatively simple transformations, the following equation:

$$d_t Y(t) - \frac{2s}{1-st}Y(t) = 0; \ s = \frac{m}{n}, \ 0 < e < t. \tag{1.3}$$

The solution to this equation, defined by use of the limiting process $e \to 0$ and the average intensity of the flow of investments $Y_0$ in the neighborhood of $t = 0$, i.e.,

$$Y(t) = \frac{Y_0}{(1-st)^2}; \ Y_0 = \frac{1}{2m(dt)}\int_{-dt}^{dt} I(h)dh, \ t \geq 0, \tag{1.4}$$

where $dt$ is a small interval of time, cardinally differs from the exponential growth (1.2).

Here, we can explicitly follow the time interval of a reasonable forecast, because, when $t$ approaches $s^{-1} = n/m$, the expression for the income (1.4) loses objective meaning. This feature is objectively inherent in a reliable model, in contrast to the dependence (1.2) that is not limited by a temporal factor, and, of course, the solution (1.4) is much more realistic.

Of interest is an extension of the outlined approach to the solution of the considered model in the case, when, e.g., $m = m_0 + m_1 t$, with an estimate of sensitivity to a small quantity $m_1$ and an analysis of corresponding behavior of the function $n(t)$ being included. One can also take into account a time lag that occurs when investments flow into the capital using a representation of the following form:

$$K(t) = K_0 + \int_0^t k(t,h) I(h) dh.$$

Let us turn to an analysis of discrete character of the initial model. Indeed, using intensities of the flows, in the case of the simplest formula for the evaluation of the definite integral, the equalities of the functions in (1.1) can only be satisfied in the following way:

$$S_c(t) = I_c(t); \; S_c(t) = mY_c(t); \; K(t) = nY_c(t), \tag{1.5}$$

where $Y_c(t) = Y(t_i); \; S_c(t) = S(t_i); \; I_c(t) = I(t_i), \; t_i \leq t \leq t_{i+1}; \; i = 0, 1, ..., n$.

As is obvious, all the functions in (1.5) are discontinuous at $t = t_i$. (Otherwise, the solution is trivial.) Thus,

$$K(t) = \int_{-T}^t I(h) dh = K_0 + \sum_{j=0}^i I(t_j), \; t_i \leq t \leq t_{i+1} \tag{1.6}$$

and, accordingly, this function is not differentiable, in the usual sense, at $t = t_i$. Its derivative can only be understood from the point of view of the theory of generalized functions, i.e., as in [2]:

$$d_t K(t) = \sum_{j=0}^i I(t_j) d(t - t_j), \; d(t) = \begin{cases} 0, t \neq 0; \\ \infty, t = 0; \end{cases} \; \int_{-\infty}^\infty d(h) dh = 1.$$

Analogously,

$$Y_c(t) = \frac{1}{n}\left[K_0 + \sum_{j=0}^i I(t_j)\right]; \; d_t Y_c(t_i) = \frac{1}{n} \sum_{j=0}^i I(t_j) d(t - t_j), \; t_i \leq t \leq t_{i+1}. \tag{1.7}$$

The functions $S_c(t)$ and $I_c(t)$ have the same structure: in other words, they are generalized functions. Thus, if one follows the methodology of [1], in reality, one uses as (1.2) the following equation:

$$d_t Y_c(t) = (m/n) Y_c(t), \tag{1.8}$$

whose solution in the class of generalized functions has the form

$$Y_c(t) = Y_{c0} e^{mt/n}. \tag{1.9}$$

However, in this case, the derivation of equation (1.8) is based on the relation

$$d_t K_c(t) = I_c(t),$$

that, from the point of view of the procedure of mathematical modeling, has no meaningful interpretation. On the contrary, the derivative $d_t K(t) = I(t)$, understood in the usual sense, objectively reflects fundamental dependence of the formation of the capital:

$$K(t) = \int_{-T}^{t} I(h) dh,$$

where $T$ is the previous period of time, as in (1.6).

In this regard, arguments of W. Kecs and P. Teodoresky [2, pp. 168-169] concerning problematical character of the use of generalized function in the derivation of differential equations seem to be rather urgent. These functions mostly serve the purpose of simplification of intermediate transformations in the process of solving the problems posed by means of continuous analysis, when the coefficients or the free terms are discontinuous.

Using (1.5) – (1.7) for $i = 0, 1, ...$, we get:

$K_0,\ Y_0 = K_0/n\ ;\ I_0 = K_0 m/n\ ;$

$K_1 = K(t_1) = K_0(1+a);\ Y_{c1} = Y_c(t_1) = K_1/n;\ I_{c1} = I_c(t_1) = K_1 m/n,\ a = m/n\ ;$

$K_2 = K_0\left[1+(1+a)a\right] = K_0(1+a+a^2);\ Y_{c2} = K_2/n;\ I_{c2} = K m/n\ ;$

$K_n = K_0 \sum_{i=0}^{n} a^i;\ Y_{cn} = K_n/n;\ I_{cn} = K_n m/n\ ,$

and, because $a < 1$ (which follows from the content of the considered model), the formula of the decreasing geometric progression yields

$$Y_{cn} = K_0 \frac{1-a^{n+1}}{n(1-a)} = Y_{c0} \frac{1-a^{n+1}}{1-a} = Y_{c0} \frac{1-(m/n)^{n+1}}{1-m/n},\ n = 0, 1, ...\ . \qquad (1.10)$$

In other words, it proves to be possible to define the solution without the use of the relation $d_t K(t) = I(t)$ from (1.1), which, generally speaking, casts doubts on the adequacy of the model [1]. Indeed, alongside with (1.10), it follows from (1.9) that

$$Y_{cn} = Y_{c0}e^{an} = Y_{c0}e^{mn/n}, \quad n = 0, 1, ..., \qquad (1.11)$$

and, accordingly, we get

$$e^{an} = (1-a^{n+1})/(1-a), \qquad (1.12)$$

which does not correspond to the approximation of the exponent in [3]. Using the analog of (1.12) with $n+1$, we arrive at the equation

$$a = \ln(1-a^{n+1})/(1-a^n),$$

which has no solution for $0 < a < 1$.

Thus, one can draw the following conclusions: the values of $Y_{cn}$, evaluated by means of formulas (1.10) and (1.11), are strikingly different. Indeed, expression (1.10) represents the solution in terms of finite difference formulation that is objectively inherent in Harrod's model [4; 5, pp. 193-199]. The solution (1.11) is erroneous because of inadequacy of the substitution of the derivative for a finite difference expression in (1.1). In other words, the exponential growth (1.2), (1.9) and (1.11) is due exclusively to incorrectness of the interpretation of the notion of infinitesimal, which is typical of a number of well-known works on macroeconomic modeling: note [6, 7].

Thus, A. Bergstrom [7] gives arguments of the following character: macroeconomics manipulates scales of decades. Against this background, a yearly income is just a small step; the exponent (1.2) smoothes out such steps, which reflects the dynamics of growth in its approximation. However, in reality, the above-mentioned "step" radically changes both the form of the differential equation and the solution to it: it is sufficient to compare (1.2), (1.8) and (1.11) with (1.3) and (1.4).

## 2. An analysis of macroeconomic models and their correction

Harrod-Domar's model of the development of the economi is presented by P. Allen [6, pp. 75-78] in the following form:

$$Y(t) = C(t) + I(t); \quad C(t) = (1-m)Y(t); \quad I(t) = n_* d_t Y(t), \qquad (2.1)$$

where $Y(t)$, $C(t)$ and $I(t)$ are intensities of the flows of income, consumption and investments, respectively; $0 < m < 1$ is a constant; the constant $n_* > 0$ has the dimension of time; $t$ is dimensional time. The differential equation of the problem and the solution to it have the form

$$d_t Y(t) = (m/n_*) Y(t); \ Y(t) = Y_0 e^{mt/n_*}, \ Y_0 = Y(0), \tag{2.2}$$

respectively.

By virtue of the fundamental dependence

$$d_t K(t) = I(t), \tag{2.3}$$

the last relation in (2.1) is equivalent to the following: $d_t K(t) = n_* d_t Y(t)$, or

$$K(t) = n_* Y(t) + K_0 - n_* Y_0, \ K_0 = K(0). \tag{2.4}$$

Eliminating the functions $Y(t)$ and $I(t)$, we arrive at the equation

$$d_t K(t) - (m/n_*) K(t) = mY_0 - (m/n_*) K_0, \tag{2.5}$$

whose solution has the form

$$K(t) = (K_0 - n_* Y_0)(e^{mt/n_*} - 1),$$

and, consequently, $K_0 = 0$, or $K_0 = n_* Y_0$ in (2.4).

In the first of these two cases, it turns out that the initial capital $K_0$ is absent, whereas the investments $I_0 = I(0)$ and the income $Y_0$ do exist, which contradicts the common sense. In the second case, equation (2.5) becomes homogeneous, and the solution to it is analogous to (2.2):

$$K(t) = n_* Y(t), \tag{2.6}$$

which also follows from (2.4). However, such dependence contradicts the use of the notion of the infinitesimal. Indeed, the intensity of income for $t = t_i$ is defined as follows:

$$Y(t_i) = \frac{1}{t_*} \int_{t_i - 0,5 t_*}^{t_i + 0,5 t_*} Y(h) dh,$$

where $t_*$ is a small interval of time. Accordingly, by (2.6), the capital is

$$K(t_i) = n \int_{t_i - 0,5 t_*}^{t_i + t_*} Y(h) dh, \ n = \frac{n_*}{t_*}. \tag{2.7}$$

In other words, $K(t_i)$ represents the value of income averaged over the interval $t_*$, and, as is obvious, the dimensionless parameter $n \to \infty$ for $t_* \to 0$.

Simultaneously, the difference between the notions of income am the intensity of income disappears at $t = t_i$. Thus, by (2.7), relation (2.6) can be considered only for a finite period $t = t_i$, and, as a consequence, it has essentially discrete character. Moreover, the use of the dimensionless time $t = t/t_*$, alongside with $n$, transforms relations (2.1) – (2.3) and (2.6) into the model [1, pp. 160-161] that, as shown in section 1, is incorrect.

An analogous situation occurs in Phillips' simplest model [6, p. 79]:

$$Z(t) = (1-m)Y(t); \; rd_t Y(t) = Z(t) - Y(t),$$

where $Z(t)$ is the flow of demand for the product; $r$ (per time unit) is the constant of a lag between demand and production. As a matter of fact, here again two functions of different dimensionality are conjugated by means of a coefficient: namely, an intensity of the flow and the rate of its change.

Phillips' general model is defined [6, pp. 81-82] by the equations

$$d_t Y(t) = l\left[I(t) - mY(t)\right]; \; d_t I(t) = k\left[nd_t Y(t) - I(t)\right], \tag{2.8}$$

where the constants and their units of measurement are the following: $k > 0$, 1/unit of time; $n > 0$, unit of time; $0 < m < 1$; $l > 0$, 1/unit of time. The problem is reduced to solving an ordinary differential equation of the second order with constant coefficients:

$$d_t^2 Y(t) + ad_t Y(t) + bY(t) = 0, \; a = k + ml - nkl; \; b = mkl, \; t \geq 0. \tag{2.9}$$

By virtue of (2.3), under the condition

$$d_t K(0) = k\left[nY(0) - K(0)\right],$$

which is an analog of the relation that follows from the solution (2.5), equation (2.8) takes the form

$$d_t Y(t) = l\left[d_t K(t) - mY(t)\right]; \; d_t K(t) = k\left[nY(t) - K(t)\right]. \tag{2.10}$$

(Note that exactly in this manner the considered model is treated by A. Bergstrom [7, pp. 40-41].)

The arguments concerning (2.6) are directly extended to the second of these equations. Accordingly, the solution to equation (2.9) does not represent an adequate reflection of the dynamics of macroeconomic development. An analogous situation also occurs for other models [6, 7]. In each case, in this or that way, there exists conjugation of the capital with an intensity of the income via a dimensional coefficient.

Following the methodology of section 1, we represent equation (2.10), relating the capital to the flows $d_t K(t)$ and $Y(t)$, in the following form:

$$K(t) = K_0 + \frac{1}{t}\int_e^t L(h)\,dh,\; L(t) = nY(t) - \frac{1}{k}d_t K(t),\; 0 < e < t, \qquad (2.11)$$

where $e$ is a small quantity. In other words, we have used a correct approach to the formation of the capital at the expense of corresponding flows of intensities by means of integration. Here, $t^{-1}$ plays the role of a proportionality factor that relates $K(t)$ to the capital accumulated by means of $Y(t)$ and $d_t K(t)$ during the period of time from $0$ to $t$.

Thus, adhering to the idea [6], put into relations of the type (2.6) and (2.10), we practically realize it on an arbitrary interval of time, including $t = e \to 0$, when the L'Hôpital rule comes into play.

From (2.11), it follows that

$$K(t) = K_0 + \frac{nk}{1+kt}\int_e^t Y(h)\,dh;\; d_t K(t) = -\frac{nk^2}{(1+kt)^2}\int_e^t Y(h)\,dh + \frac{nk}{1+kt}Y(t).$$

As a result of simple transformations, including a passage to the limit $e \to 0$, the first of relations (2.10) takes the form of the differential equation (2.9), but now its coefficients are variables:

$$a(t) = ml + \frac{2k - nkl}{1+kt};\; b(t) = \frac{2mkl}{1+kt}.$$

This equation can be represented as follows:

$$d_t^2 Y(t) + \left(a + \frac{b}{t}\right)d_t Y(t) + \frac{g}{t}Y(t) = 0,\; t \geq 1, \qquad (2.12)$$

where $t = 1 + kt$ is a dimensionless variable of time $t = 1 + kt$; the coefficients

$a = ml/k$, $a = ml/k$ and $g = 2ml/k$ are dimensionless. It can be reduced to the solution of the degenerate hypergeometric equation [8, p. 392, №2.120; pp. 428-431]. The well-known substitution [9, c. 130]

$$Y(t) = u(t)\exp\left[0.5(a - at - b\ln t)\right]$$

transforms equation (2.12) into the following:

$$d_t^2 u(t) + c(t)u(t) = 0,\ c(t) = -\frac{a^2}{4} + \frac{2g - ab}{2t} + \frac{b(2-b)}{4t^2},\ t \geq 1, \qquad (2.13)$$

from which, for

$$p = -a^2/4;\ r = (2g - ab)/2;\ s = b(2-b)/4,$$

we get

$$t^2 d_t^2 u(t) + \left(pt^2 + rt + s\right)u(t) = 0,\ t \geq 1.$$

The solution to this equation can be obtained in a closed form [8, p. 392, №2.154; pp. 547-548]. Note also the well-known substitution [9, p. 131] that reduces (2.13) to the Riccati equation $d_t u(t) + u^2(t) + c(t) = 0$; however, for a given function $c(t)$, there are no constructive methods of its solution.

To evaluate the function $Y(t)$ that satisfies (2.12) under the conditions $Y(1) = Y_1$ and $d_t Y(1) = Y_1'$, one can capitalize on the known procedure of the reduction of such a problem to a Volterra integral equation of the second kind (see, e.g., [10, pp. 16-18]). Indeed, from the notation $d_t^2 Y(t) = j(t)$ it follows that

$$Y(t) = \int_1^t (t - h) j(h) dh - (1 - t)Y_1' + Y_1,$$

and after substitution into (2.12) we get the equation

$$j(t) = \int_1^t k(t, h) j(h) dh + q(t),\ t \geq 1, \qquad (2.14)$$

where the kernel and the free term are defined by the following expressions:

$$k(t, h) = -a - b - \frac{b - h}{t};\ q(t) = -a - \frac{b - (1 - t)}{t} Y_1' + \frac{g}{t} Y_1.$$

The solution of equation (2.14) can be found with the help of the procedure of successive approximations, under a practically arbitrary choice of the initial element $j_0(t)$ [11]:

$$j_{n+1}(t) = \int_1^t k(t,h) j_n(h) dh + q(t), \quad t \geq 1; \quad n = 0, 1, \ldots . \qquad (2.15)$$

### 3. Economic-mathematical model based on price balance

Phillips' model of macroeconomic dynamics contains four sufficiently abstract parameters: the rate of reaction $k$ (an inverse of a constant lag of investments); an investment coefficient $n$ (an index of the accelerator's power); a multiplier $m$ that characterizes a part of the income directed to investment; the rate of the influence of the production output on the demand [6].

The evaluation of the intensity of the income is reduced here to solving the differential equation (2.9) that is in wide use in engineering. For example, it describes free oscillations of a mass suspended on a spring, under the condition of viscous resistance. All the parameters of such a system, including external forces, are extremely concrete and can be measured. The differential equation is derived strictly on the bases of fundamentals of mechanics [12, c. 43-49].

In this regard, one should bear in mind two points. The first one is that the spectrum of possible solutions to the above-mentioned equation is objectively insufficient for an adequate representation of macroeconomic functions. As a result, the attention of economists was attracted by equations of nonlinear theory: see, in particular, the arguments of T. Puu [13, p. 7]. However, their interpretations in categories of an objective sphere are rather problematic.

At the same time, we have shown above that Phillips' model in the interpretation [6] is incorrect. Using ideological prerequisites of this model, we have reduced the problem to the solution of a differential equation whose coefficients are time-dependent. Owing to this fact, the class of solutions has become much wider, but uncertainty in the choice of the above-mentioned parameters has remained.

The second point concerns the fact that economics does not contain laws for idealized objects that could be put in correspondence to a material point. However, economics, in its turn, has advantage over mechanics, which is embodied in the equation of the balance of financial flows. From this point of view, possibilities of mathematical modeling in economics and mechanics can be characterized as having different orientations.

Let us turn to a static system of balance equations $x = Ax + c$, or

$$x_i = \sum_{j=1}^{n} a_{ij} x_j + c_i, \, i = 1, 2, ..., n, \qquad (3.1)$$

where $x_i$ is the cost of the product of the $i$-th participant (in the case of invariable production volumes it is analogous to the price); $a_{ii}$ is the part of the cost of the product of the $i$-th participant that constitutes his income; $a_{ij}$ is the part of the cost of the product of the $j$-th participant consumed by the $i$-th participant; $c_i$ personal contribution of the $i$-th participant (including remuneration of labor, payment for materials and outside services, etc.); $t \geq 0$ is dimensional time. The coefficients $a_{ij}$ and the free terms $c_i$ are assumed to be given.

As in such a situation, except for different extraordinary factors (see below), $a_{ij}$, $c_i \geq 0$, and, obviously, the sums of the elements of each line of the matrix $A$ do not exceed unity, whereas at least one of these sums is less than unity, we have: $\|A\| < 1$ [14, pp. 329-331]. Accordingly, the quantities $x_i$ can be determined by means of successive approximations:

$$x_{s+1} = Ax_s + c, \, s = 0, 1, ... \qquad (3.2)$$

[15, pp. 120-121].

A point of principle is that the system of equations (3.1) can be attached dynamic character by setting $x = x(t)$, and

$$x_s(t) = x(t_s); \; x(t_{s+1}) = x(t_s + t_*), \tag{3.3}$$

where $t_*$ is a sufficiently small interval of time. In this regard, just the first step of the process (3.2) – (3.3) from each point $t_s = t$ of the considered interval would suffice to achieve the set goal.

Thus, there appears the relation

$$x_i(t + t_*) = \sum_{j=1}^{n} a_{ij} x_j(t) + c_i, \; t \in [0, t_*], \tag{3.4}$$

and retaining in the Taylor expansion of $x_i(t + t_*)$, say, three terms, after a transition to a dimensionless time variable $t = t/t_*$, we arrive at the following differential equation:

$$d_t^2 x_i(t) + 2 d_t x_i(t) + 2 x_i(t) = 2 \sum_{j=1}^{n} a_{ij} x_j(t) + 2 c_i, \; t \in [0, 1]. \tag{3.5}$$

By use of the initial conditions $x_i(0) = p_i$ and $d_t x_i(0) = p'_i$ (the constants $p_i$ and $p'_i$ are assumed to be given) the problem is reduced to the solution of a system of Volterra integral equations of the second kind with respect to the functions

$$j_i(t) = d_t^2 x_i(t), \tag{3.6}$$

from which it follows:

$$x_i(t) = \int_0^t (t - h) j_i(h) dh + p'_i t + p_i, \; i = 1, 2, ..., n. \tag{3.7}$$

Indeed, upon the substitution of expressions (3.6) and (3.7) into (3.5), we get

$$j_i(t) = l \sum_{j=1}^{n} \int_0^t k_{ij}(t, h) j_j(h) dh + q_i(t), \; i = 1, 2, ..., n, \tag{3.8}$$

with the parameter $l = 2$; the kernels are given by

$$k_{ij}(t, h) = \begin{cases} a_{ij}(t - h), & j \neq i; \\ (a_{ij} - 1)(t - h) - 1, & j = i; \end{cases}$$

the free terms are

$$q_i = \begin{cases} 2\left[\sum\limits_{j=1}^{n} a_{ij}\left(p'_j t + p_j\right) + c_i\right], & j \neq i; \\ 2\left[\sum\limits_{j=1}^{n} a_{ij}\left(p'_j t + p_j\right) - p'_j(1+t) - p_j + c_i\right], & j = i. \end{cases}$$

The solution to the system of equations (3.8) is obtained by means of successive approximations, i.e.,

$$j_{i,s+1}(t) = l \sum_{j=1}^{n} \int_0^t k_{ij}(t,h) j_{j,s}(h) dh + q_i(t); j_{i,0} = 0,\ s = 0, 1, \ldots; \qquad (3.9)$$

$$x_{i,s}(t) = \int_0^t (t-h) j_{i,s}(h) dh + p_{di} t + p_i,\ i = 1, 2, \ldots, n, \qquad (3.10)$$

or it can be represented as a series expansion in powers of $l$ whose terms contain sums of integrals with iterated kernels $k_{ij}(t,h)$ [11, pp. 59-61].

However, suppose that the above-mentioned requirements to $a_{ij}$ and $c_i$ are not fulfilled. For example, they can take on negative values, which reflects payment of debts, subventions, use of stocks as well as other factors of this kind. If the sum of the elements of the $j$-th column of the matrix $A$ exceeds unity, it means that the $j$-th participant sells the product to his partners at a price higher than the real value. In any case, we will consider the assumption $\|A\| < 1$ to be invalid.

In this situation, we introduce in (3.1) the notation $I - A = B$, where $I$ is a unity matrix. Now, the solution to the equation $Bx = c$ is obtained with the help of the following process of successive approximations [16, pp. 70-73]:

$$x_{s+1} = (I - rB'B)x_s + aB'c,\ s = 0, 1, \ldots$$

where $B'$ is the transpose of the matrix $B$; $0 \leq a \leq 2/\|B'B\|$.

All the arguments and transformations related to (3.2) – (3.10) still hold. Only the values of $a_{ij}$ and $c_i$ change, as, accordingly, the kernels and the free terms of equations (3.9) do. It should be noted that this fact does not influence practically the procedure of numerical realization. The same concerns retaining in (3.5) derivatives of higher order:

with increasing the length of the interval $t_*$, any a priori estimates in this sense are difficult to make.

It is not difficult to take into account in (3.1) a lag between sales of the product and production of the form $x_j(t) = x_j(t + b_j)$. The enumerated possibilities characterize a substantial advantage of the proposed approach over the solution of the considered problem in the differential form (3.5).

Another advantage of integral equations is due to a possibility to extend transformations to the case when the coefficients $a_{ij}$ and the free terms $c_i$ depend on time, which rather important in the context of further consideration. Indeed, algorithms of integration of piecewise continuous bounded functions are rather universal (see, e.g. [17]), and, from this point of view, the presence of $a_{ij} = a_{ij}(h)$ and $c_i = c_i(t)$ in (3.8) would not be of fundamental importance.

### 4. Forecasting of the development of an economic situation

Naturally, the participants of an economic system are interested in prospects of further activity. In this regard, we assume that they forecast in some way the dynamics of their mutual relations and of external demand (directly related to the cost of production) as well as the price level by the end of a considered period of time. Therefore, the functions $a_{ij}(t)$, $c_i(t)$, and the constants $x_i(1) = r_i$ are known, and we are faced with solving a boundary-value problem for the system of equations (3.5) under conditions on $x_i(t)$ for $t = 0$; $t = 1$. Using, by analogy with the previous case, (3.6), we get:

$$x_i(t) = \int_0^t (t-h) j_i(h) dh - t \int_0^1 (1-h) j_i(h) dh + (r_i - p_i) t + p_i. \qquad (4.1)$$

Upon the substitution of this expression into (3.5) the problem is reduced to the solution of a system of Fredholm integral equations of the second kind:

$$j_i(t) = l \sum_{j=1}^{n} \int_0^1 k_{ij}(t,h) j_j(h) dh + q_i(t), \; t \in [0, 1]; \; i = 1, 2, ..., n, \qquad (4.2)$$

with the parameter $l = 2$; the kernels are defined by the equations

$$k_{ij}(t,h) = \begin{cases} t(h-1)a_{ij}(h), \; t \leq h \leq 1; \\ (t-1)ha_{ij}(h), \; 0 \leq h \leq t, \; j \neq i; \end{cases}$$

$$k_{ij}(t,h) = \begin{cases} [t+1-ta_{ij}(h)](1-h), \; t < h \leq 1; \\ (t-2)ha_{ij}(h), \; 0 \leq h \leq t, \; j = i \end{cases}$$

(the second of these equations is discontinuous at the diagonal $h = t$); the free terms are

$$q_i(t) = \begin{cases} 2\left\{ \sum_{j=1}^{n} a_{ij} \left[ (r_j - p_j)t + p_j \right] + c_i \right\}, \; j \neq i; \\ 2\left\{ \sum_{j=1}^{n} (a_{ij} - 1)\left[ (r_j - p_j)t + p_j \right] - r_j + p_j + c_i \right\}, \; j = i. \end{cases}$$

For the purpose of finding the functions $j_i(t)$ that satisfy equations (4.2), a number of methods of numerical realization were developed [16]. At the same time, the system of equations (4.2) can preliminarily be reduced to a single Fredholm integral equation of the second kind [10, pp. 77-78]:

$$\Phi(t) = l \int_0^n K(t,h) \Phi(h) dh + Q(t), \; t \in [0, n], \qquad (4.3)$$

where the sought function, the free term and the kernel have the form

$$\Phi(t) = j_i(t - i + 1); \; Q(t) = q_i(t - i + 1); \; K(t,h) = k_{ij}(t - i + 1, h - j + 1),$$

$i = 1, 2, ..., n$.

A solution to this equation does exist and is unique for $l \neq l_h$, where $l_h$, $h = 1, 2, ...$ are characteristic numbers of the kernel $K(t,h)$. After the evaluation of the functions $j_i(t)$, the solution to the problem is obtained by means of substitution of these functions into (4.1).

For $l = l_h$, the homogeneous equation

$$\Phi(t) = I \int_0^n K(t,h)\Phi(h)dh; \ h = 1, 2, \ldots, t \in [0, n] \qquad (4.4)$$

has nontrivial solutions. As is obvious, an economic system should avoid such kind of critical regimes of functioning, because they are not favorable for its participants. The means for this is an increase of the efficiency of relations of mutual partnership that is implicitly related to optimization of the coefficients $a_{ij}(t)$.

Certainly, both the cost function $c_i(t)$ and the results of the activity of the participants $r_i$ can be known only approximately. Therefore, to estimate the behavior of an economic system, it is reasonable to carry out variational calculations. In other words, equation (4.3) will be solved repeatedly for a given kernel and under variations of the free term. For this reason, the following representation of the solution seems to be rather useful [11]:

$$\Phi(t) = Q(t) + I \int_0^n R(t, h, I)Q(h)dh, \ t \in [0, n], \qquad (4.5)$$

where $R(t, h, I)$ is the resolvent of the kernel $K(t, h)$. To construct the resolvent, one can use the constructive algorithm of S. G. Mikhlin [18, pp. 210-221].

From a formal point of view, the coefficients $a_{ij}(t)$ should also be varied; however, in such a case a forecast of the system's behavior can become practically unrealizable because of a large number of variants.

From this point of view, a role of such an organizing economic system as a cluster is of great importance. Indeed, its participants build up their interrelations on principles of supplying each other with reliable information and coordinate their activity on the basis of criteria of a systematic level. These facts substantially facilitate more objective determination of the function $a_{ij}(t)$.

Retaining (for better transparency of the economic situation) during the forecast period the stability of the coefficients $a_{ij}(t)$, a cluster can then redistribute incomes of the participants inside its organization. A scheme of such redistribution is preliminarily coordinated at an informal level.

Note that the outlined approach to forecasting organically matches the essence of cluster methodology: see the seminal works by M. Porter and a number of other sources [19, 20].

Let us turn to a meaningful side of the parameter $l$ in equation (4.3). At a glance, it is used only for the sake of convenience in the treatment of symbols. This point of view is justified to a certain extent, but, at the same time, one should not the origin of $l = 2$. Initially, this coefficient appeared in the procedures of the construction of the system of equations (3.8) and (4.2), and it is due to the Taylor expansion in (3.4):

$$x(t+t_*) = x(t) + \frac{1}{1!}d_t x(t) + \frac{1}{2!}d_t^2 x(t) + \frac{1}{3!}d_t^3 x(t) + \dots .$$

As we retained the first three terms, we got $l = 2$. In the case of four terms, one would have $l = 6$ and so on. However, with an increase of the interval of the Taylor series, the forecast interval $t_*$ to which the functions $a_{ij}(t)$, $c_i(t)$ and the constant $r_i$ are attached also objectively increases. Consequently, there exists an internal relation between them and the parameter $l$, which, however, cannot be expressed in functional terms. One can just state that, in reality, in (4.2) and, further, in (4.3), (4.4) we have:

$$a_{ij}(t) = a_{ij}(t, l); \; c_i(t) = c_i(t, l); \; r_i = r_i(l).$$

Accordingly, instead of (4.3), we get the equation

$$\Phi(t) = \int_0^n K(t, h, l) \Phi(t) dt + Q(t), \; t \in [0, n], \quad (4.6)$$

whose kernel depends on the parameter. As is pointed out by V. I. Smirnov, in the consideration of such equations, one can encounter substantial deviations from Fredholm's theory.

Tamarkin's theorem states that, for certain kernels $K(t, h, l)$ that depend analytically on $l$, the resolvent $R(t, h, l)$ in (4.5) does not exist for any values of this parameter [21, pp. 130-132]. In other words, equation (4.6) proves to be unsolvable (see also [10, p. 49]).

We emphasize that the above-mentioned arguments are of qualitative character because of the absence of functional dependence of $K(t, h, l)$. At the same time, the solvability of equation (4.6) and that of the initial system of equations (3.1) are, obviously, mutually related. Thus, the values $l = l_h$ in (4.4) depend in some way on characteristic numbers of the matrix $A$, whereas the insolvability of equation (4.6) is caused by the closeness to zero of its determinant $\det A$.

As regards this issue, here emerges a constructive verification of the solvability of equation (4.6), which is based on an investigation into the matrix $A$ with $a_{ij} = a_{ij}(t)$. Namely, the functions $a_{ij}(t)$ should be chosen in such a way that $\det A(t)$ should not vanish, and the matrix $A(t)$ should not be ill-defined (see, e.g., [17]) for all $t \in [0, t_*]$.

Note that ill-definedness implies, in this case, an inadequate overreaction to small perturbations both of functions fulfilled by the links and of the indices of their structural conjugation. In general, for effective functioning of such a system, including reliability of the forecast, it is desirable that the matrix $A(t)$ should satisfy the conditions of Perron-Frobenius' second theorem [22, pp. 247-248]: i.e., it should be nonnegative and indecomposable. In other words, all $a_{ij}(t) \geq 0$, and the directed graph corresponding to the matrix $A(t)$ should be strongly connected.

The latter condition means that any two vertices of the directed graph should possess a directed path that connects them [23, pp. 129-130]. However, in this case, the number of contacts in each pair of the participants proves to be rather large, which, generally speaking, is not typical of the processes of material and financial flows.

However, an advantage of the cluster exactly consists in the fact that an active exchange of experience and knowledge takes place between its participants. By definition, the cluster is characterized by a high degree of ramification of intellectual flows that, although realized on non-repayable basis, can, nevertheless, be represented in money equivalent, which thus ensures indecomposability of the matrix $A(t)$, $t \in [0, t_*]$.